\magnification 1200
\font\tenmsb=msbm10   
\font\sevenmsb=msbm7
\font\fivemsb=msbm5
\newfam\msbfam
\textfont\msbfam=\tenmsb
\scriptfont\msbfam=\sevenmsb
\scriptscriptfont\msbfam=\fivemsb

\let\nd\noindent 

\def\natural{{\rm I\kern-.18em N}}

\def\chix{{\raise.5ex\hbox{$\chi$}}}
\def\chixa{{\chix\lower.2em\hbox{$_A$}}}

\def\real{{\rm I\kern-.2em R}}
\def\integer{{\rm Z\kern-.32em Z}}
\def\complex{\kern.1em{\raise.47ex\hbox{
            $\scriptscriptstyle |$}}\kern-.40em{\rm C}}
\def\vs#1 {\vskip#1truein}
\def\hs#1 {\hskip#1truein}

\def\Month{\ifcase\number\month \relax\or January \or February \or
  March \or April \or May \or June \or July \or August \or September
  \or October \or November \or December \else \relax\fi }
\def\date{\Month \the\day, \the\year}

  \hsize=6truein        \hoffset=.25truein 
  \vsize=8.8truein      
  \pageno=1     \baselineskip=12pt
  \parskip=0 pt         \parindent=20pt
  \overfullrule=0pt     \lineskip=0pt   \lineskiplimit=0pt
  \hbadness=10000 \vbadness=10000 
\pageno=0

\footline{\ifnum\pageno=0\hss\else\hss\tenrm\folio\hss\fi}
\hbox{}
\vskip 1truein\centerline{{\bf Random Close Packing of Granular Matter}}
\vskip .2truein\centerline{by}
\vskip .2truein
\centerline{{Charles Radin
\footnote{*}{Research supported in part by NSF Grant DMS-0700120}}}
\vskip .2truein\centerline{Department of Mathematics} 
\vskip 0truein\centerline{University of  Texas} 
\vskip 0truein\centerline{Austin, TX\ \ 78712}

\vs1
\centerline{{\bf Abstract}}
\vs.1 \nd \hs0.8
We propose an interpretation of the random close packing
\vs0 \nd \hs0.8
of granular materials as a phase transition, and discuss the
\vs0 \nd \hs0.8
possibility of experimental verification.

\vs2
\centerline{October 2007}
\vs.2
\centerline{PACS Classification:\ \ 45.70.Cc, 81.05.Rm, 45.70.-n}
\vfill\eject
\hbox{} \nd
{\bf Introduction}

  The phenomenon of random close packing was popularized by Bernal [B]
  as part of his effort to model liquids. The experiments which
  clarified many of the issues were performed by Scott et al [S,SC,SK],
  and showed the following. If a large number of monodisperse hard
  spheres, for instance steel ball bearings, are gently poured into a
  container, the volume fraction will be roughly 0.61. If the
  container is repeatedly shaken vertically, this density rises to
  about 0.64, and careful protocols lead to reproducible lower and
  upper limits on the volume fraction, called, respectively, random
  loose packing ($0.608\pm .006$) and random close packing ($0.6366\pm
  0.0005$) [SK]. 
  Scott et al [SC] noted that volume fractions beyond 0.64 (up
  to approximately 0.66) could be obtained if the material was
  cyclically sheared, and that this result was accompanied by small
  crystal-like clusters of spheres. This was confirmed and explored by
  Pouliquen et al [ND] (who also performed experiments employing
  horizontal shaking [PN]), obtaining volume fractions up to 0.70,
  again accompanied by crystal-like clusters of spheres.
  (Recall that the densest possible packing of monodisperse
  spheres has volume fraction $\pi/\sqrt{18}\approx 0.74$.)
  
  In light of the above, the volume fraction 0.64 is generally
  described as the boundary between two regimes for granular matter:
  at volume fractions below 0.64 the structure of the material is
  random, while above 0.64 it has some order, as represented by the
  crystal-like clusters which appear.

  This general understanding has recently been questioned in an
  influential paper by Torquato et al [TT], in which the authors claim
  to ``have shown that the notion of RCP (random close packing) is not
  well defined mathematically''. In contrast we propose an unambiguous
  meaning to such a boundary between disordered and ordered
  states of granular matter, as a boundary between well defined
  phases, together with a mathematical model of
  traditional form and an experimental test of our interpretation.
  
 \vs.1 \nd {\bf Analysis} \vs.03 Our proposed characterization of
random close packing is motivated by properties of the hard sphere
model of classical statistical mechanics. In that model point
particles, with the usual position and momentum degrees of freedom,
interact only through a hard core: no pair may approach closer than
some fixed separation $D$, and the particles evolve dynamically
through elastic collisions of imaginary spheres of diameter $D$
surrounding their positions.  Our interest in the hard sphere model
stems from the demonstration by Alder and Wainright [AW], by molecular
dynamics simulations, that the model exhibits a first order phase
transition between a fluid, which exists at volume fractions below
$0.494\pm 0.002$, and a solid, believed to be crystalline, which
exists at volume fractions above $0.545\pm 0.002$ [HR]. Between 0.49
and 0.54 there is a mixed phase. Using the canonical ensemble we
integrate out the momentum variables and consider the ``reduced''
probability distributions on the phase space of the position variables
alone, in the infinite volume limit. They are the infinite volume
limits of the uniform distributions on packings for fixed volume
fraction. In particular the distribution $p(m)_f$ of the mixed phase
at volume fraction $f$, $0.49\le f\le 0.54$, is represented by an
average of the distributions of the pure phases: $p(m)_f
=c\,p_{0.54}+(1-c)p_{0.49}$, where $p_{0.49}$ is the distribution of
the highest density fluid, $p_{0.54}$ is the distribution of the
lowest density solid, and $0\le c\le 1 $ is such as to produce the
volume fraction $f$, namely $c=(f-0.49)/0.05$. (This is merely a
statement of the fact that distinct phases separate when coexisting in
equilibrium, each occupying a well defined volume [LL].) Note that
0.49 is the volume fraction of ``freezing''. Therefore assuming, as
generally believed, that the solid phase is crystalline (in fact face
centered cubic [W,BF]), we interpret 0.49 as the ``highest random
density'' among monodisperse spheres. (These structural features have
been confirmed not only by many computer simulations but also in
experiments with appropriate colloids [RD].)  Intuitively, at any
volume fraction above the freezing point there is a nonzero
probability of (seeing) an infinite, ordered crystal. It is the use of
the infinite volume limit, together with the probabilistic formalism,
which produces a sharp phase transition between disorder and order in
equilibrium statistical mechanics [FR].
  
  We emphasize that there are sphere packings with packing fraction
  $d> 0.49$ which might well be described as random, for instance
  packings corresponding to any metastable extension of the fluid
  branch of an
  isotherm. However the total of all such packings has
  probability zero with respect to the (infinite volume limit of the)
  uniform distribution on packings with packing fraction $d$. Our use
  of this distribution as the touchstone of relevancy is in accord
  with its common appearance in statistical physics and probability
  theory; the best justification one can give is that it is commonly
  found that practical sampling of phase space seems to occur in this
  way, in particular in the natural dynamics of matter in thermal
  equilibrium.

  In summary, the hard sphere model, and its physical realization in
  colloids, exhibits the basic ingredients needed to make sense of the
  granular phenomenon of random close packing: the volume fraction
  0.49 separates the fluid phase of random packings from the mixed
  phase in which crystalline order begins to appear. While this is not
  directly applicable to the granular matter which is our proper
  subject, it nonetheless shows that the intuitive notion of random
  close packing is not inherently inconsistent as claimed in [TT].

  We now turn to granular matter.
  The traditional hard sphere model does not include the effects of
  gravity and cannot represent the properties of granular matter, in
  particular that of random close packing. However a slightly modified
  ensemble framework has been proposed as a model for granular
  matter. Specifically, in the original proposal of Edwards et al [EO]
  one uses a uniform distribution on those static monodisperse sphere
  packings, of fixed volume fraction, which are mechanically stable
  under gravity. One can add friction to the spheres and perhaps
  other restrictions besides volume fraction; adding a condition of
  fixed pressure might be useful, though it is not clear if pressure
  is isotropic in granular materials.
  
  As is true for solids in thermal equilibrium, in order that an
  ensemble method be appropriate for ({\it nonequilibrium}) granular matter
  it is important to restrict the protocols used to produce beds of
  granular matter at fixed volume fraction (and perhaps pressure etc.)
  One feature that is necessary is that the protocol be
  ``history independent'' [NK] in that it give equivalent results
  starting with beds originally prepared in any manner. There are
  three types of protocols which claim to produce history independent
  beds of monodisperse granules with well defined volume fractions:
  vertical vibration (often called ``tapping''); fluidization followed
  by sedimentation; and cyclic shearing. These methods have produced
  history independent beds with volume fractions in the following
  ranges: 0.605 to 0.625 by vertical vibration [NK,RR]; 0.685 to 0.70
  by cyclic shearing [ND]; and 0.57 to 0.62 by
  fluidization/sedimentation [SG,SN].
  
  It is known from Schr\"oter et al [SN] that granular beds prepared
  by fluidization/sedimentation undergo a phase transition, as volume
  fraction is varied, at approximately 0.60 volume fraction, as
  measured by two different responses to shear.  Given the
  mathematical similarity between the hard sphere model and the
  granular model of Edwards on the one hand, and the experimental
  similarity between the sharp freezing transition in colloids [RD]
  and the abrupt appearance of crystalline clusters in the experiments
  of Scott et al [SC] and of Pouliquen et al [ND] on the other hand,
  we predict that history independent granular beds would show another
  phase transition: a first order phase transition, with a mixed phase
  for volume fractions between 0.64 and 0.74, again exhibited through
  the response to shear or other mechanical probe. Analogously to the
  hard sphere model, the distribution for granular matter in the mixed
  phase would be a mixture, with one component, at volume fraction
  0.74, representing a crystal, and the other, at volume fraction
  0.64, representing a disordered phase. This would give a well
  defined meaning to the phenomenon of random close packing of
  granular matter just as the freezing density defines a similar
  concept for the hard sphere model or for hard sphere colloids.

  We note the connection between this proposal and that of Kamien and
  Liu [KL], which also uses the hard sphere model to understand random
  close packing. In [KL] random close packing is associated with the
  end point of a metastable branch in the hard sphere phase diagram,
  while we use the hard sphere model only to predict behavior in a
  related but different ensemble, of packings which are mechanically
  stable under gravity, and in particular we predict a phase
  transition at volume fraction 0.64.

  History independent experimental protocols have not yet produced 
  beds with volume fraction in any interval containing the volume
  fraction of interest, 0.64; this will be necessary before our
  prediction can be checked against the behavior of granular
  matter. Alternatively it might be possible to test the prediction by
  realistic, history independent computer simulations. However it
  should be noted that fifty years of computer simulations of the
  hard sphere model have not yet been able to demonstrate the
  appearance of
  crystals at its freezing transition, so one should not be too
  optimistic that an order/disorder transition could be seen in
  simulations of granular models.

\vfill \eject
\nd
{\bf Acknowledgements}
\vs.1 
We gratefully acknowledge useful discussions with Persi Diaconis and
  Matthias Schr\"oter. We also thank the Aspen Center for Physics for
  support at the Workshop on Jamming.

\vs.5
\centerline{{\bf References}}
\vs.3 \item{[AW]} B.J. Alder and
  T.E. Wainwright, Studies in molecular dynamics II. Behavior of a
  small number of elastic spheres, J. Chem. Phys. 33 (1960) 1439-1451.
\vs.1 \item{[B]} J.D. Bernal, A geometrical approach to the
  structure of liquids, Nature No. 4655 (1959) 141-147.  
\vs.1 \item{[BF]} P.G. Bolhuis, D. Frenkel, S.-C. Muse, and D.A. Huse,
  Entropy difference between crystal phases, Nature (London) 388
  (1997) 235-236.  
\vs.1 \item{[EO]} S.F. Edwards and
  R.B.S. Oakeshott, Theory of powders, Physica A 157 (1989) 1080-1090.
\vs.1 \item{[FR]} M.E. Fisher and C. Radin, Definitions of thermodynamic 
phases and phase transitions, workshop report,
\vs0 
http://www.aimath.org/WWN/phasetransition/Defs16.pdf
\vs.1 \item{[HR]} W.G. Hoover and F.H. Ree, Melting transition and
communal entropy for hard spheres, J. Chem. Phys. 49 (1968) 3609-3617.
\vs.1 \item{[KL]} R.D. Kamien and A.J. Liu, Why is random close
packing reproducible?, Phys. Rev. Lett. 99 (2007) 155501.
\vs.1 \item{[LL]} L.D. Landau and E.M. Lifshitz, Statistical Physics,
(Pergamon Press, London, 1958), trans. E. Peierls and R.F. Peierls, \S 77.
\vs.1 \item{[ND]} M. Nicolas, P. Duru and O. Pouliquen, Compaction
  of a granular material under cyclic shear, Eur. Phys. J. E 3 (2000)
  309-314.  
\vs.1 \item{[NK]} E.R. Nowak, J.B. Knight, E. Ben-Naim,
  H.M. Jaeger and S.R. Nagel, Density fluctuations in vibrated granular
  materials, Phys. Rev. E 57 (1998) 1971-1982.  
\vs.1 \item{[PN]}
  O. Pouliquen, M. Nicolas and P.D. Weidman, Crystallization of
  non-brownian spheres under horizontal shaking, Phys. Rev. Lett. 79
  (1997) 3640-3643.  
\vs.1 \item{[RD]} M.A. Rutgers, J.H. Dunsmuir,
  J.-Z. Xue, W.B. Russel and P.M. Chaikin, Measurement of the
  hard-sphere equation of state using screened charged polystyrene
  colloids, Phys. Rev. B 53 (1996) 5043-5046.  
\vfill \eject
 \item{[RR]}
  P. Ribiere, P. Richard, P. Philippe, D. Bideau, R. Delannay, On the
  existence of stationary states during granular compaction,
  Eur. Phys. J. E 22 (2007) 249-253.  
\vs.1 \item{[S]} G.D. Scott,
  Packing of spheres, Nature (London) 188 (1960) 908-909.  
\vs.1 \item{[SC]} G.D. Scott, A.M. Charlesworth and M.K. Mak, On the random
  packing of spheres, J. Chem. Phys. 40 (1964) 611-612.  
\vs.1 \item{[SG]} M. Schr\"oter, D.I. Goldman, H.L. Swinney, Stationary
  state volume fluctuations in a granular medium, Phys. Rev. E 71
  (2005) 030301(R).  
\vs.1 \item{[SK]} G.D. Scott and D.M. Kilgour, The
  density of random close packing of spheres,
  Brit. J. Appl. Phys. (J. Phys. D) 2 (1969) 863-866.  
\vs.1 \item{[SN]} M. Schr\"oter, S. N\"agle, C. Radin and H.L. Swinney,
  Phase transition in a static granular system, Europhys. Lett. 78
  (2007) 44004.  
\vs.1 \item{[TT]} S. Torquato, T.M. Truskett and
  P.G. Debenedetti, Is random close packing of spheres well defined?,
  Phys. Rev. Lett. 84 (2000) 2064-2067.  
\vs.1 \item{[W]}
  L.V. Woodcock, Entropy difference between the face-centered cubic
  and hexagonal close-packed crystals structures, Nature (London) 385
  (1997) 141-143.
\vs.3 \nd
Electronic address: radin@math.utexas.edu

\vfill \eject
\end